\documentstyle [twocolumn,epsf]{mn}
\newcommand{\mincir}{\raise
-3.truept\hbox{\rlap{\hbox{$\sim$}}\raise4.truept\hbox{$<$}\ }}
\newcommand{\magcir}{\raise
-3.truept\hbox{\rlap{\hbox{$\sim$}}\raise4.truept\hbox{$>$}\ }}
\newcommand{\minmag}{\raise
-3.truept\hbox{\rlap{\hbox{$<$}}\raise5.truept\hbox{$<$}\ }}
\newcommand{\be}{\begin{equation}}
\newcommand{\ee}{\end{equation}}
 \newcommand{\ba}{\begin{eqnarray}}
\newcommand{\ea}{\end{eqnarray}}
\newcommand{\brr}{\begin{array}}
\newcommand{\err}{\end{array}}
\newcommand{\bc}{\begin{center}}
\newcommand{\ec}{\end{center}}

\renewcommand{\baselinestretch}{1.0}

\title[Precision Cosmology from X-ray AGN clustering]
{Precision Cosmology from X-ray AGN clustering}
\author[Spyros Basilakos \& Manolis Plionis]{Spyros Basilakos$^1$ \& Manolis
Plionis$^{2,3}$\\
\vspace{0.1cm}
$^1$ Academy of Athens, Research Center for Astronomy \& Applied
  Mathematics, Soranou Efessiou 4, 11-527, Athens, Greece\\
$^2$ Institute of Astronomy \& Astrophysics, National Observatory of Athens, 
I. Metaxa \& V. Pavlou, Palaia Penteli, 15236 Athens, Greece \\
$^3$ Instituto Nacional de Astrofisica, Optica y Electronica (INAOE)
Apartado Postal 51 y 216, 72000, Puebla, Pue., Mexico
}

\begin{document}

\maketitle

\begin{abstract}
We place tight constraints on the main cosmological parameters of 
spatially flat cosmological 
models by using the recent angular clustering results of XMM-{\it
Newton} soft (0.5-2\,keV) X-ray sources (Ebrero et al. 2009a), which
have a redshift
distribution with a median of $z\sim 1$. 
Performing a standard likelihood procedure, assuming a constant in
comoving coordinates AGN clustering evolution, the AGN bias
evolution model of Basilakos et al. (2008) and the WMAP5 value of 
$\sigma_8$, we find stringent simultaneous constraints in the 
($\Omega_{m}, $w) plane, 
with $\Omega_{m}= 0.26\pm 0.05$, w$=-0.93^{+0.11}_{-0.19}$.

\end{abstract}

\vspace{1.0cm}

\section{Introduction}
Recent studies in observational cosmology, using all
the available high quality cosmological data (Type Ia
supernovae, cosmic microwave background, baryonic acoustic
oscillations, etc), converge to
an emerging ``standard model'', which is flat and it is
described by the Friedmann equation:
$H^2(a)=8 \pi G \left[\rho_{m}(a)+\rho_{Q}(a)\right]/3$, 
with $a(t)$ the
scale factor of the universe, $\rho_{\rm m}(a)$ the density
corresponding to the sum of baryonic and cold dark matter
and an extra component
$\rho_{Q}(a)$ with negative pressure called dark energy and needed to
explain the observed accelerated cosmic expansion
(eg., Davis et al. 2007; 
Kowalski et al. 2008; Komatsu et al. 2009; Hicken et al. 2009
 and references therein).

The nature of the dark energy is currently one of the most fundamental and
difficult puzzles in physics and cosmology. Indeed, during the
last decade there has been an intense  theoretical debate among cosmologists
regarding the nature of the exotic ``dark energy''.
 Due to the absence of a physically well-motivated 
fundamental theory, various candidates 
have been proposed in the literature, among which 
a cosmological constant (constant vacuum), a time varying vacuum 
quintessence, $k-$essence, vector fields,
phantom, tachyons, Chaplygin gas and the list goes on
(eg., Ozer \& Taha 1987; 
Weinberg 1989;
Wetterich 1994; Caldewell, Dave \& Steinhardt 1998; Brax \& Martin 1999;
Peebles \& Ratra 2003; Brookfield et al. 2006; Boehmer \& Harko 2007
and references therein). The simplest type of dark energy corresponds to 
a scalar field having a self-interaction potential $V(\phi)$, with the
field energy density decreasing with a slower rate than the
matter energy density (dubbed also ``quintessence'', 
eg. Peebles \& Ratra 2003 and references therein), and the
dark energy component being described by an equation of state
$p_{Q}={\rm w}\rho_{Q}$ with w$<-1/3$. 
Note, that a redshift dependence of the equation of state parameter  
is also possible but its present functional form is  
phenomenologically based (see Chevalier \& Polarski 
2001; Linder 2003).
A particular case of ``dark energy'' is the traditional cosmological constant 
($\Lambda$) model (corresponding to w$=-1$), which appears to be
 supported by the combined analysis of the recent relevant
 observational data
(eg. Komatsu et al. 2009 and references therein).

It has been shown that the
application of the correlation function analysis on samples of
high redshift galaxies or X-ray selected AGN can be used as a useful tool 
for cosmological studies (eg. Matsubara 2004; Basilakos \& Plionis
2005; 2006).
The scope of the present study is along the same lines, 
ie., to place constraints on the ($\Omega_m, $w) parameter space of spatially flat cosmological models 
using a single cosmologically relevant experiment, ie., that of the recently derived clustering properties 
of the XMM-{\it Newton} soft (0.5-2\,keV)
X-ray point sources (Ebrero et al. 2009a). 

\section{Observed and Predicted Correlations}
\subsection{X-ray AGN Correlations}
Recently, Ebrero et al. (2009a) derived the 
angular correlation function of the soft (0.5-2\,keV) 
X-ray sources using 1063 XMM-{\it Newton} 
observations at high galactic latitudes (hereafter 2XMM).
A full description of the data reduction, source detection and flux 
estimation are presented in Mateos et al. (2008). In brief, the 
survey contains $\sim 30000$ point sources within an effective
area of $\sim 125.5$ deg$^{2}$ (for an effective flux-limit of $f_x \ge 1.4 \times
10^{-15}$ erg cm$^{-2}$ s$^{-1}$ ). 
Also, Ebrero et al. (2009a) presents the details regarding 
the angular correlation function estimation, the
various biases that should be taken into account (the amplification
bias and integral constraint), the survey luminosity
and selection functions as well as issues related to possible non-AGN
contamination, which are estimated to be $\mincir 10\%$. 

The redshift selection function of the X-ray sources, 
derived by using the soft-band luminosity function of Ebrero et al.
(2009b) which takes into account the realistic luminosity dependent density 
evolution of the X-rays sources, predicts a characteristic 
depth of $z\sim 1$.

In Figure 1, we present the X-ray AGN angular correlation function 
of the Ebrero et al. (2009a) analysis. The solid points corresponds to 
the observed angular correlation function
while the solid line represents the theoretical angular correlation
function for the best fitting cosmological model (see further
below). The insert panel of Fig.1 shows the residual, $\Delta
w(\theta)$, between observations and theory. There appears to be an
interesting sinusoidal variation with $\theta$, which merits further
investigation. Unaccounted non-linear effects, at the smallest
angular separations, could be the cause of the large $\Delta w$ values
at $\theta<80^{''}$.

\subsection{From Angular to Spatial Clustering} 
We briefly present the main points of the method used to put
cosmological constraints using the angular clustering of some extragalactic
mass-tracer. A first important important
ingredient is the use of Limber's formula 
which relates the angular, $w(\theta)$, and the spatial, $\xi(r)$,
correlation functions. Assuming flatness the 
Limber's equation can be written as:
\be 
\label{eq:wthe}
w(\theta)=2\frac{\int_{0}^{\infty} \int_{0}^{\infty} x^{4} 
\phi^{2}(x) \xi(r,z) {\rm d}x {\rm d}u}
{[\int_{0}^{\infty} x^{2} \phi(x){\rm d}x]^{2}} \;\; , 
\ee
where $\phi(x)$ is the AGN redshift selection function (the probability 
that a source at a distance $x$ is detected in the survey) and 
$x$ is the coordinate distance related to the redshift through 
$x(z)=c\int_{0}^{z} \frac{{\rm d}y}{H(y)}$
with $H(z)=H_{0}[\Omega_{m}(1+z)^{3}+\Omega_{Q}(1+z)^{3(1+{\rm w})}]^{1/2}$
and $\Omega_{Q}=1-\Omega_{m}$. 
Also, $r$ is the physical separation between two sources
having an angular separation, $\theta$, which in the
small angle approximation is given by $r \simeq 
 (1+z)^{-1} \left(u^{2}+x^{2}\theta^{2} \right)^{1/2}$ (with $u$ the
line-of-sight separation of any two sources).
The number of AGN within 
a shell $(z,z+{\rm d}z)$ is given by:
\be
\label{eq:ker}
\frac{{\rm d}N}{{\rm d}z}=\delta \omega_{s}
x^{2}(z) n_s \phi(x)\left(\frac{c}{H_{0}}\right)E^{-1}(z) \;\;,
\ee
where $\delta \omega_{s}(\simeq 125.5$ deg$^{2}$
is the effective solid angle of the survey, $E(z)=H(z)/H_{0}$ and 
$n_s$ is the comoving AGN number density at $z=0$.
The source redshift distribution ${\rm d}N/{\rm d}z$, as
already mentioned
previously, is estimated by integrating the appropriate 
Ebrero et al. (2009b) luminosity function, folding in the area
curve of the survey.

Inserting eq.(\ref{eq:ker}) into eq.(\ref{eq:wthe}), we have after
some algebra that:  
\begin{equation}
\label{eq:angu}
w(\theta)=2\frac{H_{0}}{c} \int_{0}^{\infty} 
\left(\frac{1}{N}\frac{{\rm d}N}{{\rm d}z} \right)^{2}E(z){\rm d}z 
\int_{0}^{\infty} \xi(r,z) {\rm d}u \;\;.
\end{equation} 
The spatial correlation function can be written as: $ \xi(r,z) =
 (1+z)^{-(3+\epsilon)}b^{2}(z)\xi_{\rm DM}(r)$, with $\xi_{\rm DM}(r)$ 
indicating the predicted spatial correlation function of the underlying 
matter distribution (see below) and $\epsilon$ parametrizing the type 
of AGN clustering evolution (eg. de Zotti et al. 1990) and following
K\'undic (1997) and Basilakos \& Plionis (2005; 2006) we use here the
constant in comoving coordinates clustering model, ie., $\epsilon=-1.2$.
Also, $b(z)$ is the evolution of the linear bias factor which is an 
essential ingredient for 
cold dark matter (CDM) models in order to reproduce the observed mass-tracer 
distribution (cf. Kaiser 1984; Davis et al. 1985; Bardeen et al. 1986; Benson et al. 2000).
In the current analysis we use our bias evolution model (Basilakos,
Plionis \& Ragone-Figueroa 2008), which
is based on the solution of a second order differential equation
derived by using linear perturbation theory and the 
Friedmann-Lemaitre solutions of the cosmological
field equations. Our model was initially presented
in Basilakos \& Plionis (2001; 2003) and has been recently extended 
to include the effects of halo interactions and merging (for details
see Basilakos et al. 2008).

\begin{figure}
\mbox{\epsfxsize=8.2cm \epsffile{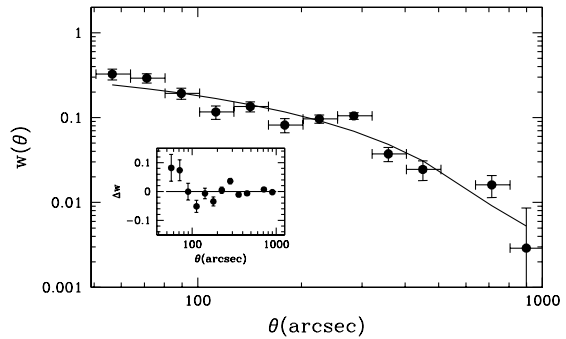}}
\caption{The two point angular correlation function
 of the soft band (Ebrero et al. 2009a). The solid line represents the 
theoretical angular correlation function for the 
best fitting ($\Omega_{m}=0.26$ and ${\rm w}=-0.93$) cosmological model.
{\em Insert Panel:} The residual angular correlation function 
between observations and theory with 2$\sigma$ uncertainties.}
\end{figure}

\subsection{Theoretically Predicted Clustering}
We estimate the theoretically predicted spatial correlation
function of the underlying matter distribution, 
$\xi_{\rm DM}(r)$, from the Fourier transform of the 
spatial power spectrum $P(k)$:
\be
\label{eq:spat1}
\xi_{\rm DM}(r)=\frac{1}{2\pi^{2}}
\int_{0}^{\infty} k^{2}P(k) 
\frac{{\rm sin}(kr)}{kr}{\rm d}k \;\;,
\ee
where $P(k)$ denotes the power of the matter fluctuations linearly 
extrapolated to the present epoch.
We consider the CDM power spectrum, $P(k)=P_{0} k^{n}T^{2}(k)$, with
$T(k)$ the CDM transfer function and
and $n\simeq 0.96$ following the 5-year WMAP results (Komatsu et al. 2009). 
In order to define the functional form of the power spectrum, 
we utilize the transfer function parameterization as in
Bardeen et al. (1986), with the approximate corrections given 
by Sugiyama (1995). The rms fluctuations of the linear density 
field on mass scale $M$ is:
\be
\sigma(M)=\left[\frac{1}{2 \pi^{2}}\int_{0}^{\infty} k^{2}P(k)
W^{2}(kR){\rm d}k \right]^{1/2} \;,
\ee
where the window function is given by:
$W(kR)=3({\rm sin}kR-kR{\rm cos}kR)/(kR)^{3}$ and
$R=(3M/ 4\pi \rho_{0})^{1/3}$.
The parameter $\rho_{0}$ denotes the mean matter density of the 
universe at the present time ($\rho_{0}=2.78 \times 10^{11}\Omega_{m}h^{2}M_{\odot}$Mpc$^{-3}$).
The normalization of the power spectrum is given by: 
$
P_{0}=2\pi^{2} \sigma_{8}^{2} \left[ \int_{0}^{\infty} T^{2}(k)
 k^{n+2} W^{2}(kR_{8}){\rm d}k \right]^{-1}  
$
where $\sigma_{8}$ is 
the rms mass fluctuation
on $R_{8}=8 h^{-1}$ Mpc scales and for which we use the WMAP5 value of
$\sigma_8\simeq 0.8$ (Komatsu et al. 2009). It is worth noting that  
we also use the non-linear corrections introduced by Peacock \& Dodds (1994).

\subsection{Cosmological Constraints}
In order to constrain the cosmological parameters 
we use, as in Basilakos \& Plionis (2005), a standard $\chi^{2}$ 
likelihood procedure and compare the measured 
XMM soft source angular correlation function (Ebrero et al. 2009a)
with the predictions of different spatially flat cosmological models.
To this end we use the likelihood estimator\footnote{Likelihoods
  are normalized to their maximum values.}, defined as:
${\cal L}_{\rm AGN}({\bf p})\propto {\rm exp}[-\chi^{2}_{\rm AGN}({\bf p})/2]$
with:
\be
\label{eq:likel}
\chi^{2}_{\rm AGN}({\bf p})=\sum_{i=1}^{n} \frac{\left[ w_{\rm th}
(\theta_{i},{\bf p})-w_{\rm obs}(\theta_{i}) \right]^{2}}
{\sigma^{2}_{i}+\sigma^{2}_{\theta_{i}}}  \;\;,
\ee 
where ${\bf p}$ is a vector containing the cosmological 
parameters that we want to estimate, $\sigma_{i}$ is the uncertainty
of the observed angular correlation function
and $\sigma_{\theta_{i}}$ corresponds to the width
of the angular separation bins.

As we have previously mentioned, 
we work within the framework of a flat cosmology 
with primordial adiabatic fluctuations and baryonic
density of $\Omega_{\rm b} h^{2}= 0.022 (\pm 0.002)$ 
(eg. Komatsu et al. 2009), while
utilizing the HST key project results of Freedman et al. (2001)
we fix the Hubble constant to $h\simeq 0.71$ (see also Komatsu et al. 2009). 
Note that since we fix, in the following analysis,
the values of both $h$ and $\Omega_b$, we do not take
into account their quite small uncertainties.

The corresponding statistical vector that we have to fit
is: ${\bf p} \equiv (\Omega_{m}, {\rm w}, M_h)$, where $M_h$ is the
host dark matter halo mass, which enters in our biasing evolution scheme (see
Basilakos, Plionis \& Ragone-Figueroa 2008). Note also that the
normalization of the power spectrum, $\sigma_8$, could have been left
as a free parameter (see Basilakos \& Plionis 2006), but in this work
we choose to use the well established WMAP5 value (see previous section).
We sample the various parameters as follows:
the matter density $\Omega_{m} \in [0.01,1]$ in steps of
0.01; the equation of state parameter ${\rm w}\in [-1.6,-0.34]$ in steps
of 0.01 and the parent dark matter halo
$M_h/10^{13}h^{-1}M_{\odot} \in [0.1,4]$ in steps of 0.01.

We find that the likelihood function of the soft X-ray sources peaks at
$\Omega_{m}=0.26\pm 0.05$, ${\rm w}=-0.93^{+0.11}_{-0.19}$ and 
$M_h=2^{+0.3}_{-0.2}\times 10^{13}\;h^{-1}M_{\odot}$, with a reduced
$\chi^{2}$ of $\sim 4$. Such a large value is caused by the small
$w(\theta)$ uncertainties in combination with the observed modulation 
(see insert panel of Fig.1). Had we used a 2$\sigma$ $w(\theta)$
uncertainty in eq.(\ref{eq:likel}) 
we would have obtained roughly the same constraints and
a reduced $\chi^{2}$ of $\sim 1$ (see upper right panel of Fig.2).

In the upper-left panel of Figure 2 we present the current constraints  
in the $(\Omega_{m},{\rm w})$ plane by marginalizing our solution over $M_h$ 
(thick solid lines). For comparison reasons we also show our previous
solutions of Basilakos \& Plionis (2005; 2006), which where based on
the shallow XMM/2dF survey ($\sim 2.3$ deg$^{2}$) 
which contains only 432 point sources 
(with an effective flux-limit of $f_x \ge 2.7 \times10^{-14}$ erg cm$^{-2}$ s$^{-1}$). 
In particular, the dotted lines correspond to a solution using
$\sigma_{8}\simeq 0.93$ (Basilakos \& Plionis 2005), while the
dashed-lines to the corresponding solution for
$\sigma_{8}\simeq 0.74$ (Basilakos \& Plionis 2006).
Comparing our present analysis with our previous results
it becomes evident that with the current high-precision X-ray AGN correlation
function of Ebrero et al. (2009a) we have achieved to place
simultaneously quite stringent constraints on both, w and $\Omega_m$.

Regarding other analyses of cosmological data, it is interesting to
note that Davis et al. (2007) using the combined analysis of the 
SNIa+BAO+CMB found $\Omega_{m}=0.27 \pm 0.04$
and ${\rm w}=-1.01 \pm 0.15$, while a similar analysis of
Kowalski et al. (2008), using a newer
SNIa compilation, provided
$\Omega_{m}=0.274 \pm 0.016$ with 
w$=-0.969 \pm 0.06$ (see also corresponding results in
Komatsu et al. 2009 and Hicken et al. 2009).
We would like to stress that despite the 
fact that we use a single cosmologically relevant experiment, ie., 
{\em the observed angular correlation function of the soft X-ray sources},
our results
coincide within $1\sigma$ 
with the results of the joint-analysis, discussed previously.
Therefore, the X-ray selected AGNs appear to be ideal tools 
for extracting cosmological information.
In order to further illustrate such a claim,
we perform further below a direct comparison between our results 
with those derived by other {\em single} cosmological data-sets.

\begin{figure}
\mbox{\epsfxsize=9cm \epsffile{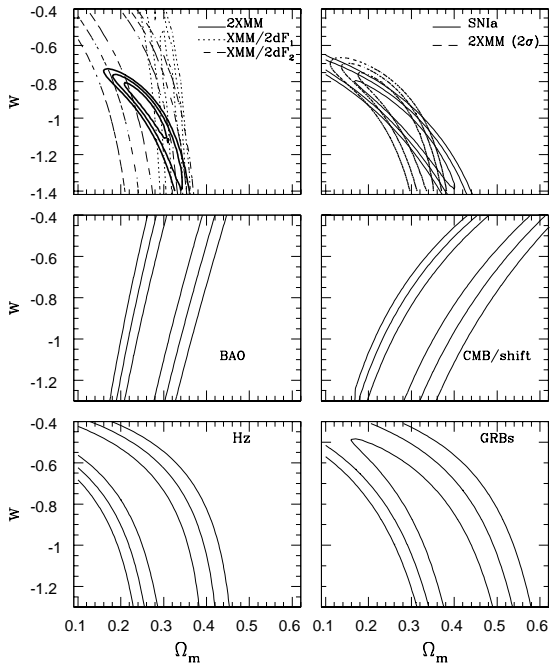}}
\caption{Likelihood contours (1$\sigma$,
2$\sigma$ and $3\sigma$) in the 
$(\Omega_{m},{\rm w})$ plane.
The upper left panel shows the results based on the current (solid
lines) and previous XMM source clustering
analyses.
In the upper right panel we
show the results based on the SNIa (solid lines) and the XMM clustering
results and a 2$\sigma$ clustering
uncertainty (dashed lines). The rest of the panels correspond to
the cosmological data indicated.
The contours 
are plotted where $-2{\rm ln}{\cal L}/{\cal L}_{\rm max}$ is equal
to 2.30, 6.16 and 11.83.}
\end{figure}

\subsection{Comparison with other Cosmological Data}
We present here a comparison between 
the ($\Omega_m, $w) solution-space provided by our analysis of 
the XMM-{\it Newton} X-ray sources 
with those derived using other cosmological data. The goal 
is to give the reader the
opportunity to appreciate the relative strength and precision of the different
methods. Therefore, we present in Figure 2 
the 1$\sigma$, 2$\sigma$ and 3$\sigma$
confidence levels in the 
$(\Omega_{m},{\rm w})$ plane, provided by the different
cosmological data presently available. These are: 
(a) the Hubble relation based on the latest 
sample of 397 supernovae of Hicken et al. (2009); (b)
the dimensionless distance to the surface of the last scattering
$R=1.71 \pm 0.019$ (Komatsu et al. 2009); (c) the baryon acoustic oscillation
distance at $z=0.35$, $A=0.469\pm 0.017$ 
(Eisenstein et al. 2005; Padmanabhan, et al. 2007); (d) the Hubble
function ($Hz$) derived directly from 9 early type galaxies at 
high redshifts
(Simon, Verde \& Jimenez 2005) and (e) the Hubble relation based on a 
sample of 69 Gamma-ray bursts (Cardone, Capozziello \& Dainotti 2009).
It is evident that the X-ray AGN clustering likelihood analysis alone
puts the most stringent
constraints on the value of the equation of state parameter. To be
fair however we must stress that in our analysis we have
{\em a priori} imposed the value of $\sigma_8$ to that of the WMAP5
(Komatsu et al. 2009),
which does not enter the other cosmological tests. However, the size of
the solution space, for any plausible value of $\sigma_8$, is as
small as for the nominal case used, the only difference would be a
tilt of the contours, as can be appreciated by the two XMM/2dF
solution spaces in the upper left panel of Fig.2.

Note, that if we increase
the uncertainty of the observed X-ray source angular correlation 
function by a factor of 2, then the resulting contours (thick dashed
lines in the right panel of figure 2) match closely those of the most
recent SNIa analysis. 
In a forthcoming paper (Plionis et al. {\em in
  preparation}) we will present details of a joint-analysis of our
X-ray selected
AGN results with that of all other cosmologically relevant data.

Lastly, we have to caution
the reader of two (reasonable, we believe) assumptions that enter 
{\em a priori} in our
analysis: (1) that the clustering evolution of X-ray selected AGN is
 constant in comoving coordinates (the effects of other
evolution models has been investigated in Basilakos \& Plionis 2005)
and (2) that the
Basilakos et al. (2008) bias evolution model is the appropriate one, 
which is supported by a comparison with N-body simulations
and available clustering data (see figures 1 and 3 of Basilakos et
al. 2008).

\section{Conclusions}
We have utilized the recent determination of the clustering properties
of high-$z$ X-ray selected AGN, identified as soft (0.5-2 keV) point
sources (Ebrero et al. 2009a),
in order to constraint the main cosmological parameters. 
We find that
the X-ray AGN clustering likelihood analysis alone, within the context
of flat cosmological models,
can place tight constrains on 
the main cosmological parameters $(\Omega_{m}, {\rm w})$,
and indeed relatively tighter than any other 
single observational method todate.
The current analysis provides a best spatially flat model with
$\Omega_{m}=0.26\pm 0.05$ and w$=-0.93^{+0.11}_{-0.19}$.

\section*{Acknowledgments} 
We are greatly thankful to Dr. J. Ebrero for comments and for providing us with an electronic
version of their clustering results and their XMM survey area-curve.
M.P. also acknowledges financial support under Mexican government CONACyT grant 2005-49878.


{\small
\begin{thebibliography}{}
\bibitem[]{}Bardeen J.M., Bond J.R., Kaiser N. \& Szalay A.S., 1986,
  ApJ, 304, 15
\bibitem[]{}Basilakos S. \& Plionis M., 2001, ApJ, 550, 522
\bibitem[]{}Basilakos S. \& Plionis M., 2003, ApJ, 593, L61
\bibitem[]{}Basilakos S. \& Plionis M., 2005, MNRAS, 360, L35
\bibitem[]{}Basilakos S. \& Plionis M., 2006, ApJ, 650, L1
\bibitem[]{}Basilakos S., Plionis M., \&, Ragone-Figueroa C., 
2008, ApJ, 678, 627
\bibitem[]{}Benson A.J., Cole S., 
Frenk S.C., Baugh M.C., \& Lacey G.C., 2000, MNRAS, 311, 793
\bibitem[]{}Brax P.,\&, Martin J., Phys. Lett. 1999, B468, 40
\bibitem[]{}Boehmer C.G., \&, Harko T., Eur. Phys. J. 2007, C50, 423
\bibitem[]{}Brookfield A.W., et al., 2006, Phys. Rev. Lett., 96, 061301 
\bibitem[]{}Caldwell R.R., Dave R., \&, Steinhardt P.J., 
1998, Phys. Rev. Lett., 80, 1582
\bibitem[]{}Cardone V.F., Capozziello S.,\&, Dainotti M.G., 2009,
{\tt (arXiv:0901.3194)} 
\bibitem[]{}Chevallier M., \&, Polarski D., Int. J. Mod. Phys. D, 2001, 10, 213
\bibitem[]{}Davis M., Efstathiou G., Frenk  C.S., \&, White S.D.M.,1985,
ApJ,292,371 
\bibitem[]{}Davis T. M., et al., 2007, ApJ, 666, 716
\bibitem[]{}de Zotti, G., Persic, M., Franceschini, A., Danese, L., 
Palumbo, G.G.C., Boldt, E.A., Marshall, F.E., 1990, ApJ, 351, 22
\bibitem[]{}Ebrero J., Mateos S., Stewart G.C., Carrera  F.J., \&,
 Watson M. G., 2009a, A\&A, in press, {\tt (arXiv0904.3024)}
\bibitem[]{}Ebrero J., et al., 2009b, A\&A, 493, 55
\bibitem[]{}Eisenstein D.J., et al., 2005, ApJ., 633, 560
\bibitem[]{}Freedman W.,L., et al., 2001, ApJ, 553, 47
\bibitem[]{}Hicken M., et al., 2009, {\tt (arXiv:0901.4804)}
\bibitem[]{}Kaiser N., 1984, ApJ, 284, L9
\bibitem[]{}Komatsu E., et al., 2009, ApJS, 180, 330
\bibitem[]{}Kowalski M., et al., 2008, ApJ, 686, 749
\bibitem[]{}K\'undic, T., 1997, ApJ, 482, 631
\bibitem[]{}Linder E. V., 2003, 90, 091301
\bibitem[]{}Mateos S., et al., 2008, A\&A, 429, 51
\bibitem[]{}Matsubara T., 2004, ApJ, 615,573
\bibitem[]{}Ozer M., \&, Taha, O., 1987, Nucl. Phys., B287, 776
\bibitem[]{}Padmanabhan N., et al., 2007, MNRAS, 378, 852
\bibitem[]{}Peebles, P.J.E.,\&  Ratra, B., Rev.Mod.Phys., 2003, 75, 559
\bibitem[]{}Peacock A.J., \&, Dodds S.J., 1994, MNRAS, 267, 1020
\bibitem[]{}Silveira V., \& Waga I., 1994, Phys. Rev. D., 64, 4890
\bibitem[]{}Simon J., Verde L., Jimenez R., 2005, Phys. Rev. D., 
71, 123001
\bibitem[]{}Sugiyama N., 1995, ApJS, 100, 281
\bibitem[]{}Weinberg S., 1989, Rev. Mod. Phys., 61, 1
\bibitem[]{}Wetterich C., A\&A, 1995, 301, 321

\end{thebibliography}
}
\end{document}